\documentclass{aastex}
\usepackage{spr-astr-addons}
\usepackage{url}
\usepackage{subfigure}

\RequirePackage{color}

\newcommand{\emaila}{antonellalucia.iannella@unisannio.it}

\begin{document}

\title{An updated comparison of the $M_{\bullet}$ vs $M_{G}\sigma^2$ relation with $M_{\bullet}$ vs $\sigma$ and the problem of the masses of galaxies}
\shorttitle{``An updated comparison of the $M_{\bullet}$ vs $M_{G}\sigma^2$ relation relation"}
\shortauthors{Iannella and Feoli}

\author{A. L. Iannella\altaffilmark{1}}
\email{antonellalucia.iannella@unisannio.it}
\email{\emaila}
\and
\author{A. Feoli\altaffilmark{1}}
\email{feoli@unisannio.it}

\altaffiltext{1}{Department of Engineering, University of Sannio, Piazza Roma 21, 82100 Benevento, Italy\\
Corresponding author: A. L. Iannella - antonellalucia.iannella@unisannio.it}

\begin{abstract}

We have studied, in a series of papers, the properties of the $M_{\bullet}$ versus $M_{G}\sigma^2$ relation and we have found that it is useful to describe the evolution of galaxies in the same way as the HR diagram does for stars and to predict the masses of Supermassive Black Holes that are difficult to be guessed using other scaling relations. In this paper, analyzing five samples of galaxies, we find that this relation has intrinsic scatter similar to  the $M_{\bullet} - \sigma$,  but follows the theoretical models much better than the $M_{\bullet} - \sigma$. Furthermore, we analyze the role of the bulge mass in the behavior of $M_{\bullet}$ versus $M_{G}\sigma^2$ relation because the difference with the $M_{\bullet} - \sigma$ is often determined by the choice of the right sample of galactic masses.

\end{abstract}


\maketitle

\section{Introduction}
An evidence of the last three decades of astrophysical observations is that almost each galaxy hosts a Supermassive Black Hole (SMBH). Another important discovery in this field of research is the existence of a correlation between the mass of SMBHs and the properties of the host galaxies, such as the velocity dispersion \citep{fer2000,geb2000,tre2002}, the bulge luminosity or mass \citep{kor1995, van1999, ric1998, mag1998, mar2001, mer2001, lao2001, wan2002, geb2003, mar2003, har2004, gul2009}, the galaxy light concentration \citep{gra2001}, the effective radius \citep{mar2003}, the Sersic index \citep{gra2005, gra2007}, the kinetic energy \citep{feo2005,feo2007}, the inner core radius \citep{lau2007}, the gravitational binding energy and gravitational potential \citep{all2007}, the momentum parameter
\citep{soker10}, the number of globular clusters
\citep{burkert10,snyder11}, the spiral arm pitch angle \citep{sei2008,ber2013}.\\
Hence there is a co-evolution of galaxies and their central SMBHs that can be described and studied in the light of the scaling relations found. The aim of our paper is neither to compare all these relations to discover the best one, nor to study all their interesting  applications or  predictions. We focus our analysis  only on  one of them.

Fifteen years ago \citeauthor{feo2005} \citeyearpar{feo2005} proposed a new correlation between the mass of a supermassive black hole and the kinetic energy of the host galaxy. The main results of this line of research that have been found during this period of time are summarizable in this way:
1) there is no doubt that the correlation exists. It has been tested with a lot of different samples and fitting methods \citep{ben2013,feo2007,feo2009,man2012};
2) there is no doubt that the relation is very competitive with all the others to fit the experimental data, in particular its intrinsic scatter is very low \citep{sag2016} just like the more popular $M_{\bullet}-\sigma$ relation and the $M_{\bullet}\propto M_{G}^{(1/2)}\sigma^2$ proposed by Hopkins (\citeyear{hop2007a});
3) the relation can be very useful to understand the evolution of galaxies, just like the HR diagram is for the evolution of stars \citep{feo2009} and allows good predictions of the masses of some black holes that do not follow the $M_{\bullet}-\sigma$ \citep{ben2013} as well as of the behavior of AGN \citep{man2012}.

In this paper we want to study other two aspects of the relation that have not been deepened before:

 1) the role of the masses of galaxies that makes the relation different from the $M_{\bullet}-\sigma$;

  2) the correspondence of the scaling laws with possible theoretical models.

  Since in recent papers there is a trend  to reduce all the scaling laws to only one, considered as the most important (``the $M_{\bullet}-\sigma$ relation is the optimum universal relation...The Fundamental Plane and the $M_{\bullet}-\sigma$ relation together constitute a basis that can define other scaling relations applicable to galaxies of all types." -- \citealt*{van2016}), we will measure the performance of our relation compared to the $M_{\bullet}-\sigma$ used as a reference standard.

\section{Samples}
For the relation $M_{\bullet}\propto M_{G}\sigma^2$, it is important to have a sample of the masses as homogeneous as possible, since it is precisely the mass of the galaxy that makes the difference with respect to the relation $M_{\bullet}-\sigma$. Therefore, we have identified three possible homogeneous databases of the masses of galaxies in the recent literature \citep{cap2013,sag2016,van2016} and used them to compose five samples of objects that form the starting point of our statistical analysis. First of all we underline that we have not taken the entire database of Saglia because we consider useless to repeat their so detailed analysis  performed in \citet{sag2016}, so we assumed it appropriate to consider their data starting from a subsample. We have avoided to choose directly which galaxies to consider and which to neglect and we have adopted the selection  made by \citet{den2019}, who have discarded 26 galaxies from the Saglia's sample, retaining only 71.
Furthermore, we know that \citet{van2016} does not report explicitly the masses of galaxies, but following Cappellari's suggestion \citep{cap2006}, we have calculated the dynamical masses in this way:
\begin{equation}
M_{dyn}=\frac{5R_e \sigma^2}{G}
\end{equation}
where $R_e$ is the effective radius of  host spheroidal component, $\sigma$ the velocity dispersion of the host galaxy and $G$ the gravitational constant.

After these two specifications, the samples we considered are as follows:\\
\begin{itemize}
  \item[$\bullet$] The \textit{1st Sample} is composed of 47 early-type galaxies, obtained by intersecting the data of \citet{cap2013} for masses of early-type galaxies and velocity dispersions, and \citet{van2016} for the corresponding masses of supermassive black holes. The only exception is for the mass of the black hole within the galaxy NGC4486, for which the most recent estimated value is considered \citep{eve2019}.
      NGC 4429 was excluded from the intersection, as its morphological classification is doubtful.
      \vspace{10pt}
  \item[$\bullet$] The \textit{2nd Sample} is obtained from \citet{van2016} considering, among his large collection of data, only the 181 galaxies with the mass of BH measured with a relative error on $log(M_{\bullet})$ not greater than or equal to 1. Furthermore, we have excluded from this subset NGC404, because its mass is too low for a SMBH and NGC4486b that ``deviates strongly from every correlation involving its black hole mass" \citet{sag2016}. In the end, we have preferred to exclude also NGC221, NGC1277, NGC1316, NGC5845 and UGC1841, reaching the final number of 174 galaxies for our \textit{2nd Sample}.
      \vspace{10pt}
  \item[$\bullet$] The \textit{3rd Sample} is nothing else but the subsample of the previous one, obtained considering only the 108 early-type galaxies.
      \vspace{10pt}
  \item[$\bullet$] Our \textit{4th Sample} consists of 71 elements, carried out from \citet{den2019}, of which we consider the velocity dispersions and masses of SMBHs associated with galaxies, while the respective masses of galaxies $M_{G}$ are taken from \citet{sag2016}, where they are denoted by the symbol $M_{Bu}$.
      \vspace{10pt}
  \item[$\bullet$] Finally, the \textit{5th Sample} is composed by the same  71 galaxies of the previous one, but all the data  were entirely extrapolated from \citet{sag2016}.

\end{itemize}
The files used for the fit containing the data of the five samples are available at the link\\ http://people.ding.unisannio.it/feoli/IF2020.zip

\section{Two-parameter Fit}

We have considered the five samples described in the previous Section and we have looked  for  the best fit line using the linear regression routines LINMIX\_ERR and MPFITEXY that work well when there are experimental errors on both variables. Then, we have also reported the ordinary least squares fit of Mathematica as a test in which we neglect the errors on the experimental data.

 The linear regression routine LINMIX\_ERR  allows to determine the slope, the normalization, and the intrinsic scatter $\varepsilon _{0}$ (which is that part of the variance
that cannot be attributed to specific causes -- \citealt*{nov2006}) of the relation
\begin{equation}
\log (M_{\bullet })=b + m\log(x) + \varepsilon _{0}.
\end{equation}%
In our case the two relations that we want to study have $x= \sigma$ or $x = M_G \sigma^2/c^2$, where $c$ is the speed of light.
LINMIX\_ERR is a Bayesian fitting method already used by us and several authors in other papers, hence we avoid to describe it  and invite to read the reference \citep{kel2007} for details.

The frequentist statistical approach can be followed starting from
the routine  FITEXY \citep{pre1992}, that minimizes the $\chi^{2}$:
\begin{equation}
\chi ^{2}=\sum_{i=1}^{N}\frac{(y_{i}-b-mx_{i})^{2}}{(\Delta
y_{i})^{2}+m^{2}(\Delta x_{i})^{2}}.
\end{equation}
To obtain the most efficient and unbiased estimate of the slope, it is necessary to introduce  the  intrinsic scatter $\varepsilon _{0}$. It can be done using MPFITEXY, an evolution of  FITEXY \citep{tre2002}.  When the reduced $\chi _{\mathrm{red}}^{2}=\chi ^{2}/(N-2)$ of the fit is not equal to 1,  MPFITEXY  automatically normalizes $\chi _{\mathrm{red}}^{2}$
including $\varepsilon _{0}$ in the equation (3):
\begin{equation}
\chi^2_{\mathrm{red}} = \frac{1}{N-2} \sum_{i=1}^{N} \frac{(y_i -b-m x_i)^2}{
(\Delta y_i)^2 + \varepsilon_{0}^2+ m^2 (\Delta x_i)^2}.
\end{equation}
\\

Finally, we also used a standard routine of \textit{Mathematica} that performs  an ordinary least squares fit, without considering the errors.
We have analyzed the five samples of galaxies respectively, and we have obtained with MPFITEXY, LINMIX\_ERR and with \textit{Mathematica} the values shown in Table 1.

We have also included in the table the Pearson linear correlation coefficient calculated with the formula
\begin{equation}
R=\frac{\sum_{i=1}^{n}(x_{i}-\bar{x})(y_{i}-\bar{y})}{\sqrt{%
\sum_{i=1}^{n}(x_{i}-\bar{x})^{2}}\sqrt{\sum_{i=1}^{n}(y_{i}-\bar{y})^{2}}}%
\;.
\medskip
\end{equation}%

From an inspection of the tables it is evident that the slope of $M_{\bullet}- M_{G}\sigma^2$ relation is near the unity for the first three samples, while it drastically changes for the last two samples where it decreases to $m = 0.72$. On the other side the slope of $M_{\bullet}-\sigma$ relation oscillates around $m = 5$ for all the five samples.
Considering the values of the intrinsic scatter and of the correlation coefficient, we observe that the two relations are on the same level for the first sample,  while for the third, fourth and fifth samples the $M_{\bullet}- M_{G}\sigma^2$ relation has slightly better values than the $M_{\bullet}-\sigma$ and the opposite occurs for the second sample.

Since van den Bosch's sample has a sufficient number of spirals, we can plot in Figure 1 the relation $M_{\bullet}- M_{G}\sigma^2$, highlighting the various morphological types with different colors. Lenticular galaxies are excluded from the plot, as they are uniformly scattered. The distribution of galaxies in the figure is such that there is a separation between the elliptical galaxies placed in the upper right part of the plot and the barred lenticulars, spirals and barred spirals placed in the lower left part, as already shown in previous papers on this subject \citep{ben2013,feo2009}.

\begin{figure*}[ht!]
\centerline{\includegraphics{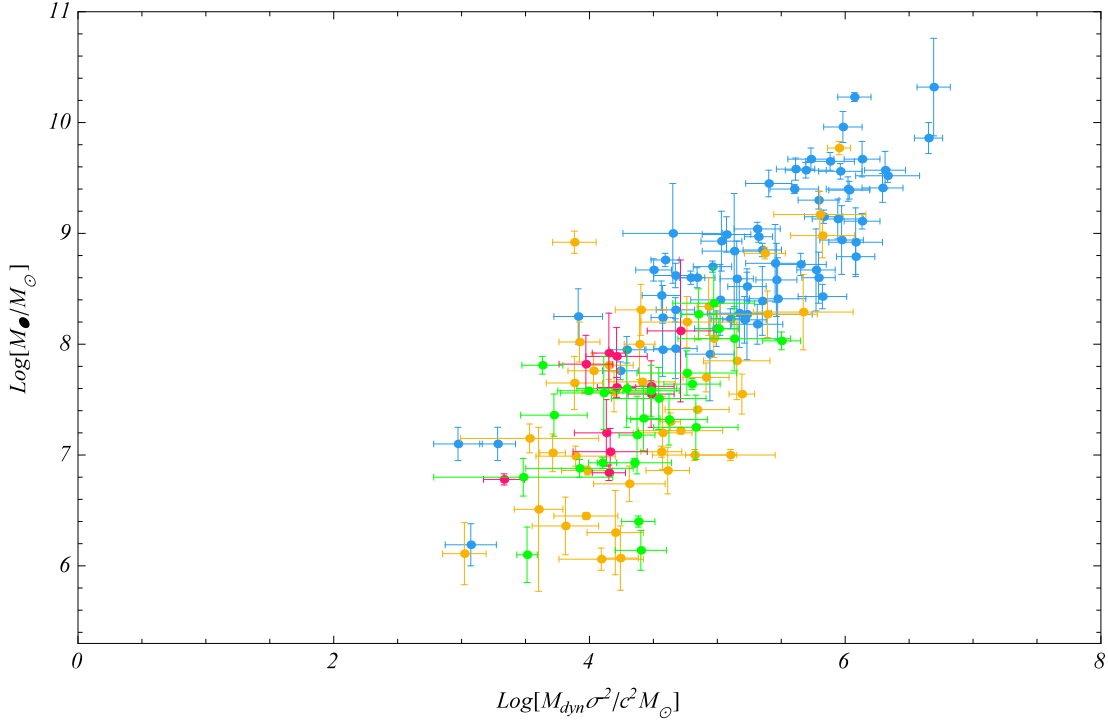}}
\caption{2nd Sample is plotted with different colors for different morphological type. It is possible to see in blue the elliptical galaxies, in fuchsia the lenticular barred or with a weak bar, in yellow the spirals and in green the intermediate and barred spirals.}
\end{figure*}

\begin{table*}[!h]
\small
\begin{center}
\textbf{Table 1a.} Two-parameter Fits for Cappellari's Sample\\
\smallskip
\begin{tabular}{@{}cccccccccccc@{}}
\tableline
\tableline
\noalign{\smallskip}
\multicolumn{12}{c}{\textsl{\textbf{1st Sample: Cappellari}}} \\
\noalign{\smallskip}
&& \multicolumn{4}{c}{\textbf{$Log\left(M_{\bullet}\right)-$ $Log\left(\displaystyle M_{JAM}\sigma^2 \over \displaystyle c^2 \right) $}}&&\multicolumn{4}{c}{\textbf{$Log\left(M_{\bullet}\right)-Log\left(\sigma\right)$}}\\
& & $m$ & $b$ & $\epsilon_0$ & $R$ && & $m$ & $b$ & $\epsilon_0$ & $R$ \\
\tableline
\noalign{\smallskip}
\vspace*{4pt}\textsl{LINMIX\_ERR} &&0.99$\pm 0.10$& 3.78$\pm 0.46$&  0.39$\pm 0.06$ &--&&& 5.15$\pm 0.51$ & -3.34$\pm 1.15$ & 0.38$\pm 0.06$ &--\\
\vspace*{4pt}\textsl{MPFITEXY} && 0.99$\pm 0.09$& 3.76$\pm 0.42$&    0.35  &--&&&
5.20$\pm 0.46$ & -3.44$\pm 1.05$ & 0.35 &--\\
\textsl{Mathematica} && 0.96$\pm 0.09$& 3.88$\pm 0.41$&   0.413 &  0.845&&& 4.88$\pm 0.46$ & -2.75$\pm 1.03$ & 0.413 & 0.845\\
\tableline
\end{tabular}
\end{center}
\smallskip

\textsl{Note:} $m$ and $b$ are the slope and the intercept of the linear relation respectively, $\epsilon_0$ is the intrinsic scatter of the relation and R the linear correlation coefficient shown for $M_{\bullet}-$ $\displaystyle M_{JAM}\sigma^2 \over \displaystyle c^2$ and $M_{\bullet}-\sigma$.
\end{table*}

\section{One-parameter Fit}
After the standard analysis contained in the previous section, we want to propose a different point of view that consists in the assumption for the two relations of a fixed slope that can be better explained   by a theoretical  model. We choose to test for the $M_{\bullet}- \sigma$ relation the slopes $m = 4$ and $m = 5$, while for the $M_{\bullet}- M_{G}\sigma^2$ relation the values $m = 0.75$ and $m = 1$ (\citeauthor{feo2014} \citeyear{feo2014} and references therein). Furthermore, in recent papers, the statistical analysis of the scaling relations is done using more and more sophisticated machinery. Our different point of view consists also in coming back to a basic approach. We choose a very simple
method for fitting the considered data sets that is the ordinary least squares. It comes down, for the one parameter case, to  calculate, with the following exact formulas, the intercept $b$ \citep{feo2007} of the best fit line:

\begin{equation}
b =\frac{\sum_{i=1}^N \bigl(\frac {y_i-m x_i}{m^2(\triangle x_i)^{2}+(\triangle y_i)^{2}}\bigr )}{\sum_{i=1}^N \bigl(\frac {1}{m^2( \triangle x_i)^{2}+(\triangle y_i)^{2}}\bigr )}
\end{equation}

and its uncertainty

\begin{equation}
(\bar{\triangle b})^2 =\frac{1}{\sum_{i=1}^N \bigl(\frac {1}{m^2(\triangle x_i)^{2}+(\triangle y_i)^{2}}\bigr )}
\end{equation}

In one-parameter fit the reduced $\chi^2$ (4) must be divided by $N-1$ and not by $N-2$.
For the uncertainty on the intercept we have used a  slightly different formula from (7) that is used to calculate ``the standard error of the weighted mean (scale corrected)". Indeed,  when the value of $\chi^2_{\mathrm{red}}$ is too large, the right uncertainty $(\triangle b)^2$ is obtained multiplying $(\bar{\triangle b})^2$ given by (7) for the $\chi^2_{\mathrm{red}}$. The corresponding results for the five samples are displayed in Table 2.

In all the samples  for the $M_{\bullet}- \sigma$ relation, the slope $m =5$  is better than the $m=4$, as we expected. For the $M_{\bullet}- M_{G}\sigma^2$ relation, we have $m =1$ for the first three samples and $m= 0.75$ for the last two samples, as we had already argued in the previous section. The difference is that for the first three samples the $M_{\bullet}- M_{G}\sigma^2$ relation gives significantly better values of $\chi^2_{\mathrm{red}}$ with respect to the $M_{\bullet}- \sigma$ relation and the same occurs for the last two samples, comparing the result of $m=0.75$ for $M_{\bullet}- M_{G}\sigma^2$ with the other relation having $m= 5$.

\begin{table*}[!t]
\small
\begin{center}
\textbf{Table 1b.} Two-parameter Fits for van den Bosch's Samples\\

\smallskip
\begin{tabular}{@{}cccccccccccc@{}}
\tableline
\tableline
\noalign{\smallskip}
\multicolumn{12}{c}{\textsl{\textbf{2nd Sample: van den Bosch\_174}}} \vspace*{2pt}\\

\noalign{\smallskip}
&& \multicolumn{4}{c}{\textbf{$Log\left(M_{\bullet}\right)-$ $Log\left(\displaystyle M_{dyn}\sigma^2 \over \displaystyle c^2\right)$}}&&
\multicolumn{4}{c}{\textbf{$Log\left(M_{\bullet}\right)-Log\left(\sigma\right)$}}\\

& & $m$ & $b$ & $\epsilon_0$ & $R$ && & $m$ & $b$ & $\epsilon_0$ & $R$ \\
\tableline
\noalign{\smallskip}
\vspace*{4pt} \textsl{LINMIX\_ERR} &&1.01$\pm 0.06$& 3.27$\pm 0.28$&    0.52$\pm 0.04$ &  --&&& 5.06$\pm 0.26$&-3.30$\pm 0.60$&0.49$\pm 0.03$&--\\
\vspace*{4pt}
\textsl{MPFITEXY} && 1.02$\pm 0.05$& 3.21$\pm 0.26$&    0.49  &--&&  &  5.10$\pm 0.25$&-3.39$\pm 0.56$& 0.47&--\\
\textsl{Mathematica} && 0.95$\pm 0.05$& 3.52$\pm 0.26$&    0.569  &0.807&&  &  4.89$\pm 0.24$&-2.91$\pm 0.55$& 0.526&0.838\\
\tableline

\end{tabular}
\end{center}

\end{table*}

\begin{table*}[!t]
\small
\smallskip
\begin{center}
\begin{tabular}{@{}cccccccccccc@{}}
\tableline
\tableline
\noalign{\smallskip}
\multicolumn{12}{c}{\textsl{\textbf{3rd Sample: van den Bosch\_108}}}
\vspace*{2pt}\\
\noalign{\smallskip}
&& \multicolumn{4}{c}{\textbf{$Log\left(M_{\bullet}\right)-$ $Log\left(\displaystyle M_{dyn}\sigma^2 \over \displaystyle c^2\right)$}}&&
\multicolumn{4}{c}{\textbf{$Log\left(M_{\bullet}\right)-Log\left(\sigma\right)$}}\\
& & $m$ & $b$ & $\epsilon_0$ & $R$ && & $m$ & $b$ & $\epsilon_0$ & $R$ \\
\tableline
\noalign{\smallskip}
\vspace*{4pt} \textsl{LINMIX\_ERR} &&0.91$\pm 0.05$& 3.95$\pm 0.26$&    0.37$\pm 0.04$ &  --&&& 4.93$\pm 0.28$&-2.89$\pm 0.64$&0.38$\pm 0.03$&--\\
\vspace*{4pt}
\textsl{MPFITEXY} && 0.92$\pm 0.05$& 3.93$\pm 0.24$&    0.34  &--&&  &  4.94$\pm 0.26$&-2.92$\pm 0.60$& 0.35&--\\
\textsl{Mathematica} && 0.86$\pm 0.05$& 4.17$\pm 0.24$&    0.423  &0.871&&  &  4.73$\pm 0.25$&-2.44$\pm 0.59$& 0.416&0.875\\

\tableline
\end{tabular}
\end{center}
\smallskip

\textsl{Note:} $m$ and $b$ are the slope and the intercept of the linear relation respectively, $\epsilon_0$ is the intrinsic scatter of the relation and R the linear correlation coefficient shown for $M_{\bullet}-$ $\displaystyle M_{dyn}\sigma^2 \over \displaystyle c^2$ and $M_{\bullet}-\sigma$.
\end{table*}

\begin{table*}[!t]
\small
\begin{center}
\textbf{Table 1c.} Two-parameter Fits for 4th and 5th Samples\\
\smallskip
\begin{tabular}{@{}cccccccccccc@{}}
\tableline
\tableline
\noalign{\smallskip}
\multicolumn{12}{c}{\textsl{\textbf{4th Sample: de Nicola - Saglia}}} \vspace*{2pt}\\

\noalign{\smallskip}
&& \multicolumn{4}{c}{\textbf{$Log\left(M_{\bullet}\right)-$ $Log\left(\displaystyle M_{Bu}\sigma^2 \over \displaystyle c^2\right)$}}&&
\multicolumn{4}{c}{\textbf{$Log\left(M_{\bullet}\right)-Log\left(\sigma\right)$}}\\

& & $m$ & $b$ & $\epsilon_0$ & $R$ && & $m$ & $b$ & $\epsilon_0$ & $R$ \\
\tableline
\noalign{\smallskip}
\vspace*{4pt} \textsl{LINMIX\_ERR} &&0.72$\pm 0.04$& 5.19$\pm 0.17$&    0.35$\pm 0.04$ &  --&&& 4.98$\pm 0.28$&-3.09$\pm 0.64$&0.37$\pm 0.04$&--\\
\vspace*{4pt}
\textsl{MPFITEXY} && 0.72$\pm 0.04$& 5.19$\pm 0.17$&    0.34  &--&&  &  4.99$\pm 0.26$&-3.11$\pm 0.61$& 0.35&--\\
\textsl{Mathematica} && 0.72$\pm 0.04$& 5.19$\pm 0.17$&    0.386  &0.919&&  &  4.92$\pm 0.27$&-2.94$\pm 0.62$& 0.405& 0.911\\
\tableline

\end{tabular}
\end{center}

\end{table*}

\begin{table*}[!h]
\small
\smallskip
\begin{center}
\begin{tabular}{@{}cccccccccccc@{}}
\tableline
\tableline
\noalign{\smallskip}
\multicolumn{12}{c}{\textsl{\textbf{5th Sample: Saglia}}}\\
\vspace*{2pt}\\
\noalign{\smallskip}
&& \multicolumn{4}{c}{\textbf{$Log\left(M_{\bullet}\right)-$ $Log\left(\displaystyle M_{Bu}\sigma^2 \over \displaystyle c^2\right)$}}&&
\multicolumn{4}{c}{\textbf{$Log\left(M_{\bullet}\right)-Log\left(\sigma\right)$}}\\
& & $m$ & $b$ & $\epsilon_0$ & $R$ && & $m$ & $b$ & $\epsilon_0$ & $R$ \\
\hline
\noalign{\smallskip}
\vspace*{4pt} \textsl{LINMIX\_ERR} &&0.73$\pm 0.04$& 5.16$\pm 0.18$&    0.36$\pm 0.04$ &  --&&& 5.04$\pm 0.27$&-3.22$\pm 0.64$&0.37$\pm 0.04$&--\\
\vspace*{4pt}
\textsl{MPFITEXY} && 0.73$\pm 0.04$& 5.16$\pm 0.17$&    0.34  &--&&  &  5.05$\pm 0.27$&-3.24$\pm 0.62$& 0.36&--\\
\textsl{Mathematica} && 0.72$\pm 0.04$& 5.17$\pm 0.17$&    0.392  &0.919&&  &  4.97$\pm 0.27$&-3.05$\pm 0.62$& 0.407&0.912\\
\tableline
\end{tabular}
\end{center}
\smallskip

\textsl{Note:} $m$ and $b$ are the slope and the intercept of the linear relation respectively, $\epsilon_0$ is the intrinsic scatter of the relation and R the linear correlation coefficient shown for $M_{\bullet}-$ $\displaystyle M_{Bu}\sigma^2 \over \displaystyle c^2$ and $M_{\bullet}-\sigma$.
\end{table*}

\begin{table*}[!h]
\small
\begin{center}
\textbf{Table 2a.} One-parameter Fits for Cappellari's Sample\\
\smallskip
\begin{tabular}{@{}ccccccccccc@{}}
\tableline
\tableline
\noalign{\smallskip}
\multicolumn{11}{c}{\textsl{\textbf{1st Sample: Cappellari}}}\\
\noalign{\smallskip}
&& \multicolumn{3}{c}{\textbf{$Log\left(M_{\bullet}\right)-$ $Log\left(\displaystyle M_{JAM}\sigma^2 \over \displaystyle c^2\right)$}}&&&&\multicolumn{3}{c}{\textbf{$Log\left(M_{\bullet}\right)-Log\left(\sigma\right)$}}\\
& & $m=0.75$ && $m=1$ & & &&  $m=4$ && $m=5$ \\
\tableline
\noalign{\smallskip}
\vspace*{4pt} \textsl{$b\pm \Delta b$} &&4.89$\pm 0.06$&&3.71$\pm 0.06$ &&&& -0.69$\pm 0.06$&&-2.97$\pm 0.06$\\
\vspace*{4pt}
\textsl{$\chi ^2$} &&432.70&&257.45 &&&& 436.51&&268.17\\
\textsl{$\chi_{red}^2$} &&9.41&& 5.60&&&& 9.49&&5.83 \vspace*{2pt}\\
\tableline

\end{tabular}
\end{center}
\smallskip

\textsl{Note:} $b$ is the intercept with its uncertainty, the $\chi^2$ and $\chi^2_{\mathrm{red}}$ shown for $M_{\bullet}-$ $\displaystyle M_{JAM}\sigma^2 \over \displaystyle c^2$ and $M_{\bullet}-\sigma$ with the fixed slopes.

\end{table*}

\begin{table*}[!h]
\small
\begin{center}
\textbf{Table 2b.} One-parameter Fits for van den Bosch's Samples\\
\smallskip
\begin{tabular}{@{}ccccccccccccccccccc@{}}
\tableline
\tableline
\noalign{\smallskip}
\multicolumn{6}{c}{\textsl{\textbf{2nd Sample: van den Bosch\_174}}}&&\multicolumn{6}{c}{\textsl{\textbf{3rd Sample: van den Bosch\_108}}}&&&& \vspace*{2pt}\\
\noalign{\smallskip}
& \multicolumn{2}{c}{\textbf{$Log\left(M_{\bullet}\right)-$ $Log\left(\displaystyle M_{dyn}\sigma^2 \over \displaystyle c^2 \right) $}}&\multicolumn{2}{c}{\textbf{$Log\left(M_{\bullet}\right)-Log\left(\sigma\right)$}}
&&\multicolumn{2}{c}{\textbf{$Log\left(M_{\bullet}\right)-$ $Log\left(\displaystyle M_{dyn}\sigma^2 \over \displaystyle c^2 \right) $}}&\multicolumn{2}{c}{\textbf{$Log\left(M_{\bullet}\right)-Log\left(\sigma\right)$}}\\
& $m=0.75$ & $m=1$ &    $m=4$ & $m=5$&
& $m=0.75$ & $m=1$ &    $m=4$ & $m=5$ \\
\tableline
\noalign{\smallskip}
\vspace*{4pt} $b\pm \Delta b$ &4.64$\pm 0.05$& 3.41$\pm 0.04$&
-0.81$\pm 0.04$ &-3.11$\pm 0.04$&&
4.86$\pm 0.04$& 3.58$\pm 0.04$ & -0.68$\pm 0.04$ & -3.03$\pm 0.04$\\
\vspace*{4pt}
$\chi ^2$ & 1881.61 & 1123.93 & 2296.00 &1631.02&&
652.18 & 425.33 & 957.50 & 659.58\\
$\chi_{red}^2$& 10.88& 6.50 & 13.27 & 9.43&&
 6.10 & 3.98 & 8.95 & 6.16\\
\tableline

\end{tabular}
\end{center}
\smallskip

\textsl{Note:} $b$ is the intercept with its uncertainty, the $\chi^2$ and $\chi^2_{\mathrm{red}}$ shown for $M_{\bullet}-$ $\displaystyle M_{dyn}\sigma^2 \over \displaystyle c^2$ and $M_{\bullet}-\sigma$ with the fixed slopes.

\end{table*}

\begin{table*}[!h]
\small
\begin{center}
\textbf{Table 2c.} One-parameter Fits for 4th and 5th Samples\\
\smallskip
\begin{tabular}{@{}ccccccccccccccccccccc@{}}
\tableline
\tableline
\noalign{\smallskip}
\multicolumn{6}{c}{\textsl{\textbf{4th Sample: de Nicola - Saglia}}}&&\multicolumn{6}{c}{\textsl{\textbf{5th Sample: Saglia}}}&&&&&&&& \vspace*{2pt}\\
\noalign{\smallskip}
& \multicolumn{2}{c}{\textbf{$Log\left(M_{\bullet}\right)-$ $Log\left(\displaystyle M_{Bu}\sigma^2 \over \displaystyle c^2\right)$}}
&\multicolumn{2}{c}{\textbf{$Log\left(M_{\bullet}\right)-Log\left(\sigma\right)$}}
&&\multicolumn{2}{c}{\textbf{$Log\left(M_{\bullet}\right)-$ $Log\left(\displaystyle M_{Bu}\sigma^2 \over \displaystyle c^2\right)$}}
&\multicolumn{2}{c}{\textbf{$Log\left(M_{\bullet}\right)-Log\left(\sigma\right)$}}\\
& $m=0.75$ & $m=1$ &    $m=4$ & $m=5$&
& $m=0.75$ & $m=1$ &    $m=4$ & $m=5$ \\
\tableline
\noalign{\smallskip}
\vspace*{4pt} $b\pm \Delta b$ &5.06$\pm 0.04$ & 4.00$\pm 0.07$ & -0.84$\pm 0.05$ & -3.16$\pm 0.05$&&
5.08$\pm 0.04$ & 4.03$\pm 0.07$ & -0.83$\pm 0.05$ & -3.15$\pm 0.05$\\
\vspace*{4pt}
$\chi ^2$ & 579.97 & 1113.63 & 975.74 &668.69&&
592.93 & 1094.53 & 1056.27 & 707.51\\
$\chi_{red}^2$& 8.29& 15.91 & 13.94 & 9.55&&
 8.47 & 15.64 & 15.09 & 10.11\\
\tableline

\end{tabular}
\end{center}
\smallskip

\textsl{Note:} $b$ is the intercept with its uncertainty, the $\chi^2$ and $\chi^2_{\mathrm{red}}$ shown for $M_{\bullet}-$ $\displaystyle M_{Bu}\sigma^2 \over \displaystyle c^2$ and $M_{\bullet}-\sigma$ with the fixed slopes.

\end{table*}

\begin{figure*}[!t]
\centering%
\subfigure[]
{\includegraphics[width=79mm]{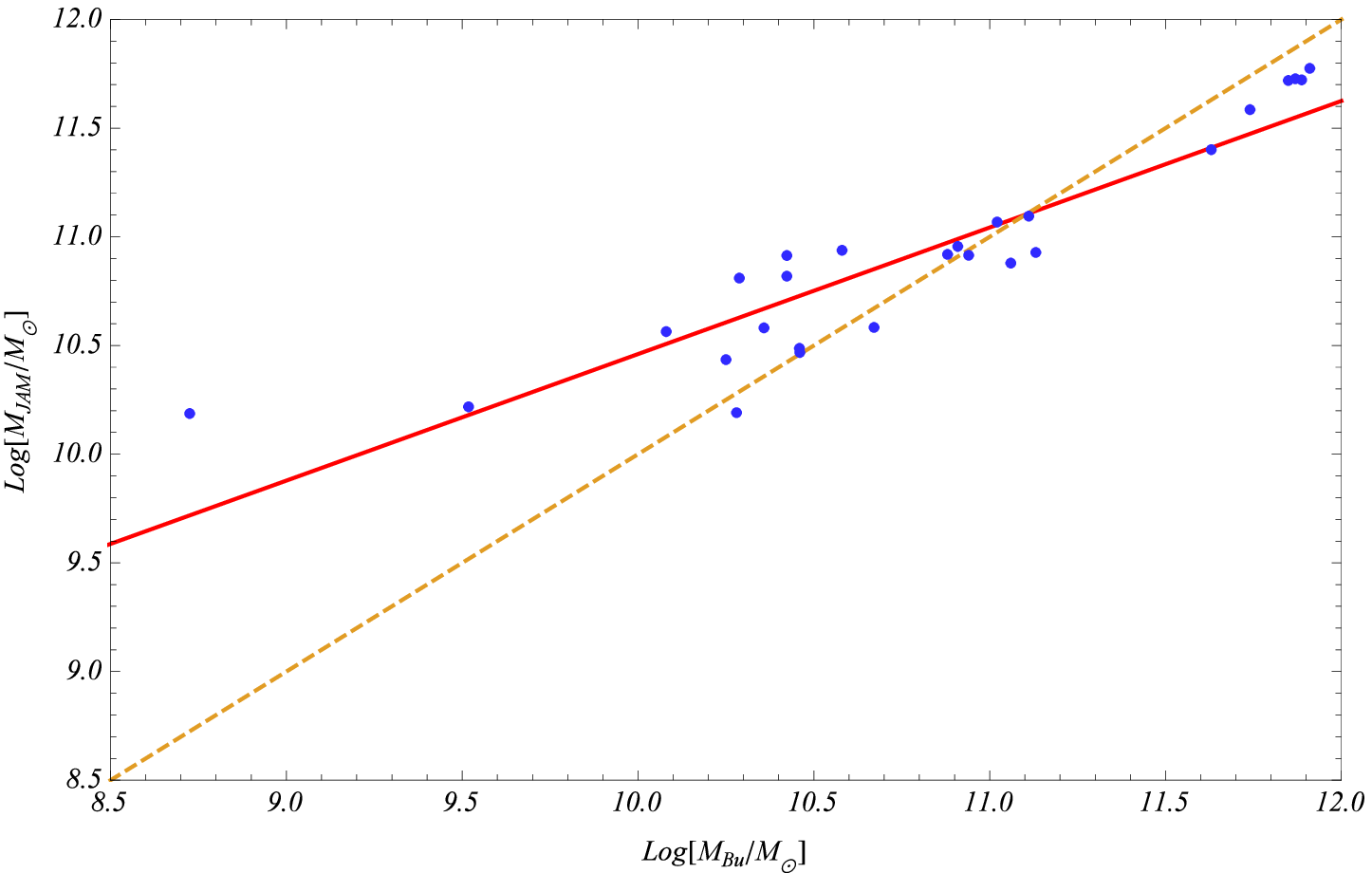}}\qquad\qquad
\subfigure[]
{\includegraphics[width=79mm]{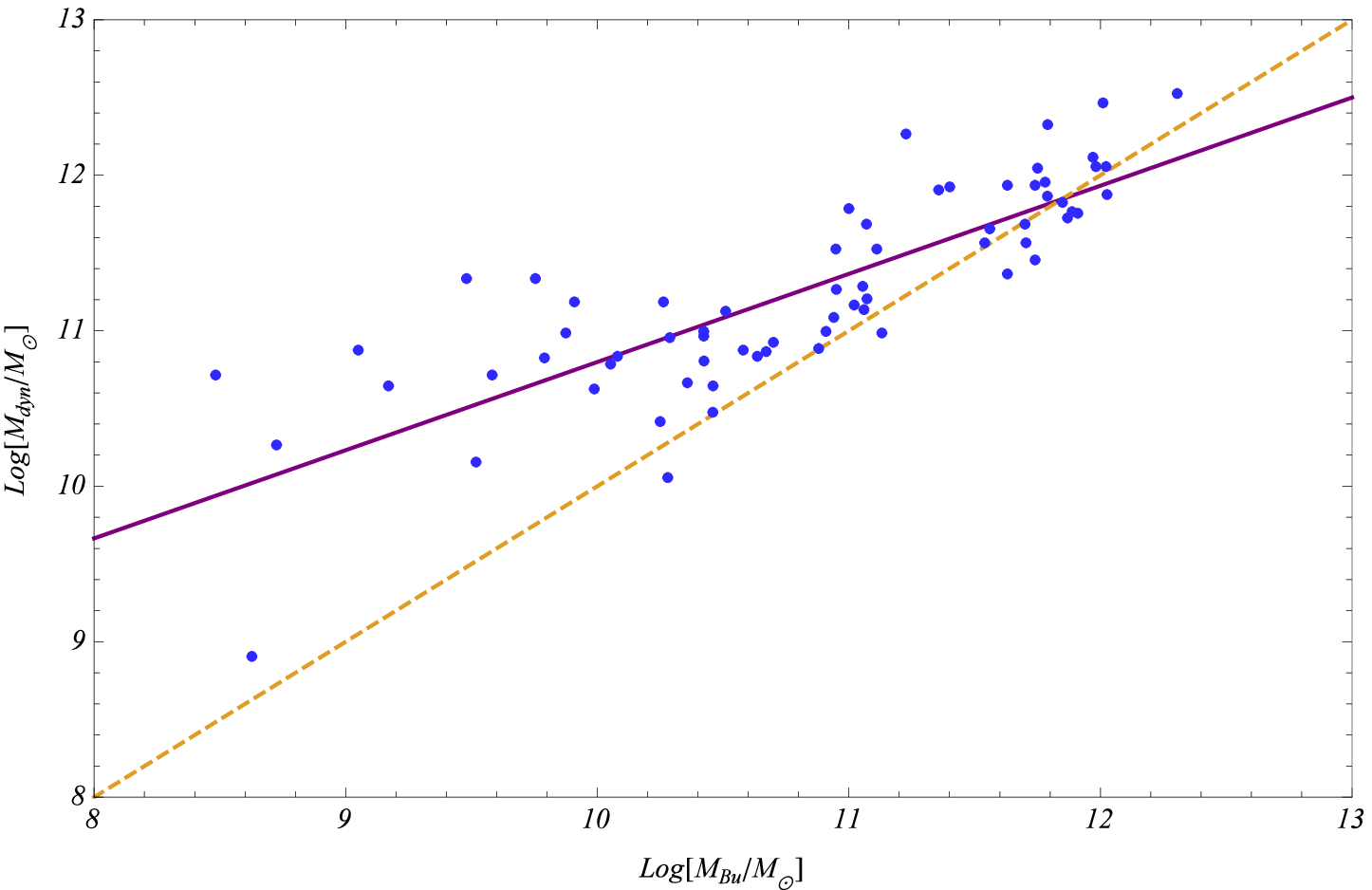}}\qquad\qquad
\subfigure[]
{\includegraphics[width=79mm]{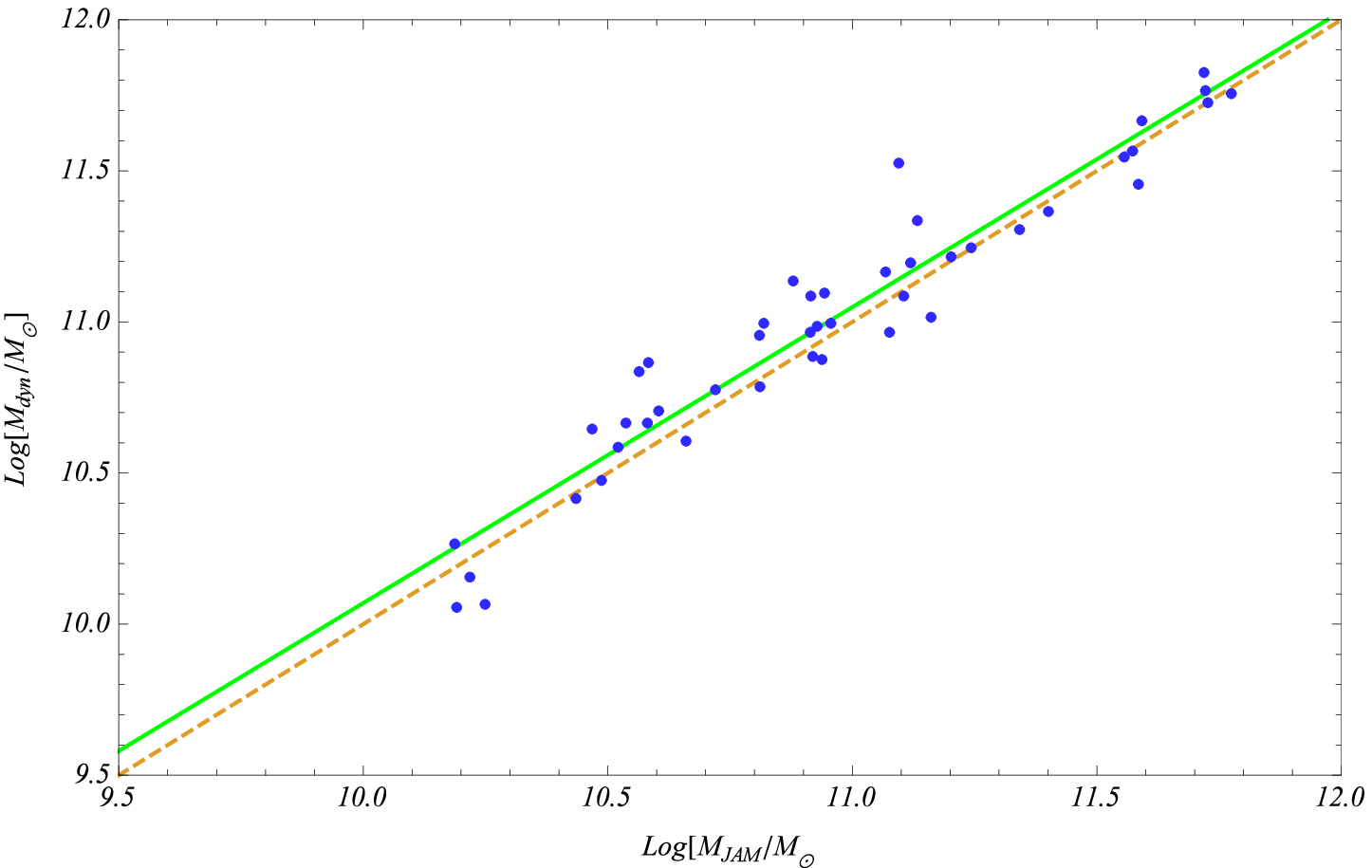}}
\caption{The comparison of the masses of the galaxies extracted from different databases. Considering only the galaxies common to both samples, we have plotted: (a) Cappellari-Saglia, (b) van den Bosch-Saglia and (c) van den Bosch-Cappellari, where the solid line indicates the best fit, while the dashed line  indicates what is expected if the masses of both samples have the same values. \label{fig:sottofigure}}
\end{figure*}

\section{The Problem of the Masses of Galaxies}
The estimation of the mass of a galaxy can be done by using different methods (see for example Appendix B of \citeauthor{feo2009}\citeyear{feo2009}) and each author, according with the aim of his paper, can also choose to calculate the total mass of the galaxy  or only one of its components: bulge, disk, dark matter halo. The consequence is that the different estimation of galactic masses is reflected in the resulting slopes and intercepts of the scaling relations.
Analyzing the various fits for the $M_{\bullet}\propto M_{G}\sigma^2$ relation, it has been noticed that there was a strong difference between the best fit angular coefficient obtained for the fourth and fifth samples  using the Saglia's  masses $(M_{Bu})$ and those predicted in all the other cases $(M_{JAM}$ and $M_{dyn})$. What we can deduce about the different slope is that:
\begin{itemize}
\item[$\bullet$] it does not depend on the estimate of errors. In fact we have used three different fit programs, in two of which the errors have been considered in the fits, while in the third \textit{Mathematica} the errors have not been considered. The results were almost unchanged, so we can deduce that this is not what affects the final results.
\vspace{10pt}
\item[$\bullet$] It does not depend on the choice of the method to fit, because for each sample the three routines find almost the same slope within the limits of uncertainty.
\vspace{10pt}
\item[$\bullet$] It does not depend on the morphological type of galaxies, since, both analyzing the Cappellari's sample, where we have only early-type galaxies, and analyzing the van den Bosch's sample, where 65 spirals are present, we have still obtained  similar values. So we can say that considering the spirals or not, the final result does not change.
\end{itemize}
For the considerations just explained, we are led to think that this difference depends on the estimation of the masses, i.e. masses estimated with different methods lead to a different slope.
In order to confirm this point of view, we have made three graphics (Figure 2) formed by taking only the values of the masses of galaxies that are in common between the two  samples considered for  each plot (Saglia $\rightarrow M_{Bu}$, Cappellari $\rightarrow M_{JAM}$ and van den Bosch $\rightarrow M_{dyn}$). What we noticed is that Cappellari-van den Bosch's masses follow the bisector $(y=x)$ of the graphic, while this does not happen in the other two plots, where the masses of the galaxies of Saglia are present.
The best fit lines shown in the three plots were derived using \textit{Mathematica} and lead to the following relations:
\begin{equation}
M_{dyn} \propto M_{Bu}^{(0.567)} \simeq (M_{Bu})^{1/2}
\end{equation}

\begin{equation}
M_{JAM} \propto M_{Bu}^{(0.582)} \simeq (M_{Bu})^{1/2}
\end{equation}

\begin{equation}
M_{dyn} \propto M_{JAM}^{(0.979)} \simeq M_{JAM}
\end{equation}
\\
Thanks to (8) and (9), the Feoli and Mele's relation $M_{\bullet}\propto M_{G}\sigma^2$ (\citeyear{feo2005}) can be confused with the one proposed by \citeauthor{hop2007a} (\citeyear{hop2007a}, \citeyear{hop2007b}), $M_{\bullet}\propto M_G ^{1/2}\sigma^2$, because $M_{\bullet}\propto M_{JAM}\sigma^2$ and  $M_{\bullet}\propto M_{dyn}\sigma^2$ are equivalent to
$M_{\bullet}\propto M_{Bu}^{1/2}\sigma^2$.

\section{Conclusions}
We have analyzed the behavior of the $M_{\bullet}\propto M_{G}\sigma^2$ relation using five samples of galaxies taken by three different sources. The result is that the relation works as well as the $M_{\bullet}-\sigma$. Furthermore, we have  fixed the slope of the two relations to understand their concordance with possible theoretical models. In fact, relations of the kind $M_{\bullet}\propto (M_{G}\sigma^2)^{0.79}$ or $M_{\bullet} \propto \sigma^{4.63}$ have poor physical meaning. The one - parameter fit (at least with the samples used in this paper) shows that the matching of the $M_{\bullet}\propto M_{G}\sigma^2$ relation with the slopes $m=1$ or $m= 0.75$ is significantly better than the matching of $M_{\bullet}-\sigma$  with the slope $m=4$ or $m=5$. Finally, the difference in the slope found using different samples is often due to the estimation of bulge masses. We have found a not trivial difference for  the mass of the same object in different databases and this fact can induce a confusion between the $M_{\bullet}\propto M_{G}\sigma^2$ relation and the $M_{\bullet}\propto M_{G}^{1/2}\sigma^2$ proposed by \citeauthor{hop2007a} (\citeyear{hop2007a}).

\section*{\footnotesize Acknowledgements}
{\footnotesize The authors thank Carmela Galdi for allowing us to use some computer facilities of her laboratory for our analysis.\\
This research was partially supported by FAR fund of the University of Sannio.}

\end{document}